\documentclass[10pt, conference]{IEEEtran}
\usepackage{subfigure}
\usepackage{setspace}
\usepackage{amsmath}
\usepackage{amssymb}
\usepackage{amsfonts}
\usepackage{amscd}
\usepackage{mathrsfs}
\usepackage[final]{graphicx}
\usepackage{graphicx}
\usepackage{psfrag}
\usepackage{color}
\usepackage{url}
\usepackage{bm}

\newtheorem{theorem}{Theorem}

\newtheorem{definition}{Definition}
\newtheorem{proposition}{Proposition}

\newtheorem{case}{Case}

\begin{document}

\title{STBCs with Reduced Sphere Decoding Complexity for Two-User MIMO-MAC}

\author{
\authorblockN{J. Harshan}
\authorblockA{Dept of ECE, Indian Institute of Science \\
Bangalore 560012, India\\
Email:harshan@ece.iisc.ernet.in\\
}
\and
\authorblockN{B. Sundar Rajan}
\authorblockA{Dept of ECE, Indian Institute of Science \\
Bangalore 560012, India\\
Email:bsrajan@ece.iisc.ernet.in\\
}
}

\maketitle

\begin{abstract}
In this paper, Space-Time Block Codes (STBCs) with reduced Sphere Decoding Complexity (SDC) are constructed for two-user Multiple-Input Multiple-Output (MIMO) fading multiple access channels. In this set-up, both the users employ identical STBCs and the destination performs sphere decoding for the symbols of the two users. First, we identify the positions of the zeros in the $\textbf{R}$ matrix arising out of the Q-R decomposition of the lattice generator such that (i) the worst case SDC (WSDC) and (ii) the average SDC (ASDC) are reduced. Then, a set of necessary and sufficient conditions on the lattice generator is provided such that the $\textbf{R}$ matrix has zeros at the identified positions.  Subsequently, explicit constructions of STBCs which results in the reduced ASDC are presented. The rate (in complex symbols per channel use) of the proposed designs is at most $2/N_{t}$ where $N_{t}$ denotes the number of transmit antennas for each user. We also show that the class of STBCs from complex orthogonal designs (other than the Alamouti design) reduce the WSDC but not the ASDC.
\end{abstract}

\begin{keywords}
MIMO, multiple access channels, sphere decoder, low decoding complexity codes.
\end{keywords}

\section{Introduction and Preliminaries}
\label{sec1}
\indent Two-user Gaussian multiple access channels (MAC) with finite complex input alphabets and continuous output have been studied in \cite{HaR} wherein the impact of the rotation between the alphabets of the two users on the capacity region has been investigated. Coding schemes for the above channel model has also been proposed in \cite{HaR4}. Recently, the idea of rotation between the alphabets of the two users is extended to MIMO fading MAC in \cite{HaR2} wherein Space-Time Block Code (STBC) pairs with low Maximum Likelihood (ML) decoding complexity and information-losslessness property are proposed for a two-user MIMO (Multiple-Input Multiple-Output) fading MAC (See Fig. \ref{mimo_mac_model} for the two-user MIMO-MAC model). In the channel model considered in \cite{HaR2}, it is assumed that the destination has the perfect knowledge of Channel State Information (CSI) and the two users have the perfect knowledge of only the phase components of their channels to the destination (referred as CSIT-P). When CSIT-P is not available, STBCs proposed in \cite{HaR2} are not applicable. For some earlier works on space-time coding for MIMO-MAC, we refer the reader to \cite{GaB} and the references therein. Note that STBCs with minimum ML decoding complexity have been well studied in the literature for co-located MIMO channels \cite{TJC, Xl} and distributed MIMO channels as well \cite{YiK, HaR3}.

In this paper, a two-user MIMO fading MAC with $N_{t}$ antennas at both the users and $N_{r}$ antennas at the destination is considered with the assumption that the destination has the perfect knowledge of CSI and the users do not have CSI. For such a set-up, STBC pairs are proposed such that the sphere decoding \cite{ViB}, \cite{DHC} complexity is reduced. The contributions of the paper may be summarized as below :
\begin{itemize}
\item In a two-user MIMO fading MAC, when both the users employ identical STBCs from linear complex designs \cite{HaH} and the destination performs sphere decoding for the symbols of the two users, we identify a class of complex designs which results in a special class of lattice generators called row-column (RC) monomial lattice generators. (Definition \ref{def_2} in Section \ref{sec2}). Employing Q-R decomposition on RC monomial lattice generators, we identify the positions of the zeros in the $\textbf{R}$ matrix such that the worst case sphere decoding complexity (WSDC) and/or the average sphere decoding complexity (ASDC) is reduced (Definition \ref{def_3} and Definition \ref{def_4}). Further, a set of necessary and sufficient conditions on the RC monomial lattice generators is provided such that the $\textbf{R}$ matrix has zeros at the identified positions (Theorem \ref{thm_2} in Section \ref{sec3}).
\item We explicitly construct STBCs which reduce the ASDC. The rate of the proposed STBCs in complex symbols per channel use per user is at most $\frac{2}{N_{t}}$.  We also show that STBCs from the class of complex orthogonal designs (other than the Alamouti design) only reduce the WSDC (but not the ASDC). (Section \ref{sec4}).
\end{itemize}
\textit{Notations:} Throughout the paper, boldface letters and capital boldface letters are used to represent vectors and matrices respectively. For a complex matrix $\textbf{X}$, the matrices $\textbf{X}^*$, $\textbf{X}^T$,  $\textbf{X}^{H}$, $|\textbf{X}|$, $\mbox{Re}(\textbf{X})$ and $\mbox{Im}(\textbf{X})$ denote, respectively, the conjugate, transpose, conjugate transpose, determinant, real part and imaginary part of $\textbf{X}$. For any matrix $\textbf{X}$, $\textbf{X}_{c}(j)$ denotes the $j$-th column of $\textbf{X}$ and $[\textbf{X}]_{i,j}$ denotes the element in the $i$-th row and the $j$-th column of $\textbf{X}$. The tensor product of the matrix $\textbf{X}$ with itself $r$ times is represented by $\textbf{X}^{\otimes^r}$. Cardinality of the set $\mathcal{S}$ is denoted by $|\mathcal{S}|$. Absolute value of a complex number $x$ is denoted by $|x|$ and $E \left[x\right]$ denotes the expectation of the random variable $x$. A circularly symmetric complex Gaussian random vector, $\textbf{x}$ with mean $\bm{\mu}$ and covariance matrix $\mathbf{\Gamma}$ is denoted by $\textbf{x} \sim \mathcal{CN} \left(\bm{\mu}, \mathbf{\Gamma} \right)$. The inner product of two vectors $\textbf{x}, \textbf{y} \in \mathbb{R}^{T \times 1}$ is denoted by $\langle \textbf{x}, \textbf{y} \rangle$. The set of all real diagonal matrices is denoted by $\mathcal{D}$. For any complex vector $\textbf{x} \in \mathbb{C}^{k \times 1}$, $\vec{\textbf{x}}$ is given by
\begin{equation*}
\vec{\textbf{x}} = \left[ \mbox{Re}(\textbf{x})^{T} ~\mbox{Im}(\textbf{x})^{T} \right]^{T} \in \mathbb{R}^{2k \times 1}.	
\end{equation*}

The remaining content of the paper is organized as follows: In Section \ref{sec2}, the MIMO-MAC model considered in this paper is described. In Section \ref{sec3}, conditions on the lattice generator are presented such that the WSDC and/or ASDC are reduced. In Section \ref{sec4}, STBCs with reduced WSDC and reduced ASDC are presented. Section \ref{sec5} constitutes conclusion and some directions for possible further work.\\
\section{Channel model}
\label{sec2}
\begin{figure}
\centering
\includegraphics[width=3.5in]{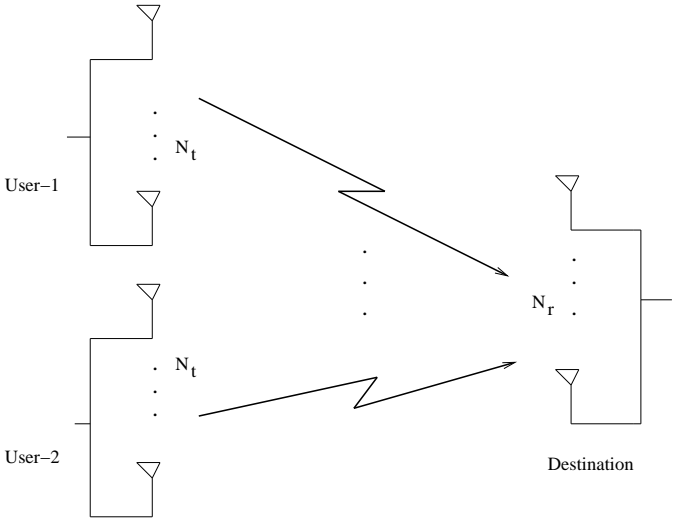}
\caption{Two-user MIMO-MAC model} 
\label{mimo_mac_model}
\end{figure}
The two-user MIMO-MAC model considered in this paper (see Fig. \ref{mimo_mac_model}) consists of two users each equipped with $N_{t}$ antennas and a destination equipped with $N_{r}$ antennas. The MIMO channels from User-1 to the destination and User-2 to the destination are respectively denoted by $\textbf{H}^{(1)} \in \mathbb{C}^{N_{t} \times N_{r}}$ and $\textbf{H}^{(2)} \in \mathbb{C}^{N_{t} \times N_{r}}$ where $[\textbf{H}^{(1)}]_{i,j}$, $[\textbf{H}^{(2)}]_{i,j} \sim \mathcal{CN} \left(0, 1\right)$ $\forall ~i = 1 \mbox{ to } N_{t} \mbox{ and } j = 1 \mbox{ to } N_{r}$. The two MIMO channels are assumed to be quasi-static with a coherence time of at least $T$ channel uses. Let $\mathcal{C}_{1}$ and $\mathcal{C}_{2}$ represent STBCs of dimension $T \times N_{t}$ employed by User-1 and User-2 respectively. If $\textbf{X}_{1} \in \mathcal{C}_{1}$ and $\textbf{X}_{2} \in \mathcal{C}_{2}$ are the codeword matrices transmitted from User-1 and User-2 simultaneously, the received matrix $\textbf{Y} \in \mathbb{C}^{T \times N_{r}}$ at the destination is given by,
\begin{equation}
\label{mimo_mac_channel}
\textbf{Y} = \sqrt{\frac{\rho}{2N_{t}}}\textbf{X}_{1}\textbf{H}^{(1)} + \sqrt{\frac{\rho}{2N_{t}}}\textbf{X}_{2}\textbf{H}^{(2)} + \textbf{N},
\end{equation}
where $\textbf{N} \in \mathbb{C}^{T \times N_{r}}$ is the additive noise at the destination such that each component of $\textbf{N}$ is distributed as $\mathcal{CN} \left(0, 1\right)$. In this model, we have assumed equal average power constraint for both the users. Assuming $E\left[\left[\textbf{X}_{1}\right]_{t, i}\left[\textbf{X}_{1}\right]^{*}_{t, i}\right] = 1$ and $E\left[\left[\textbf{X}_{2}\right]_{t, i}\left[\textbf{X}_{2}\right]^{*}_{t, i}\right] = 1$ for all $t = 1$ to $T$ and $i = 1$ to $N_{t}$, the average receive signal to noise ratio (SNR) at the destination is $\rho$. Throughout the paper, we assume perfect knowledge of both $\textbf{H}^{(1)}$ and $\textbf{H}^{(2)}$ at the destination for every codeword use.\\
\indent We construct STBC pairs $(\mathcal{C}_{1}, \mathcal{C}_{2})$ such that the sphere-decoding complexity at the destination is reduced. The STBC pair $(\mathcal{C}_{1}, \mathcal{C}_{2})$ is specified by presenting a complex design pair $(\textbf{X}_{1}, \textbf{X}_{2})$ and a complex signal set pair $(\mathcal{S}_{1}, \mathcal{S}_{2})$ such that $\mathcal{C}_{1}$ and $\mathcal{C}_{2}$ are generated by making the complex variables of $\textbf{X}_{1}$ and $\textbf{X}_{2}$ take values from the signal sets $\mathcal{S}_{1}$ and $\mathcal{S}_{2}$ respectively. We only consider the class of linear designs for $\textbf{X}_{1}$ and $\textbf{X}_{2}$ \cite{HaH}. In particular, identical designs are employed for both the users and hence in order to distinguish the two designs, the complex variables of the design for User-1 and User-2 are denoted by $\left\lbrace x_{11}, x_{12} \cdots x_{1k}\right\rbrace$ and $\left\lbrace x_{21}, x_{22} \cdots x_{2k}\right\rbrace$ respectively where $k$ denotes the number of complex variables in the design. Since the designs are linear, they can be represented as shown below,
\begin{equation*}
\textbf{X}_{1} = \left[\textbf{A}_{1}\textbf{x}_{1} + \textbf{B}_{1}\textbf{x}_{1}^{*} ~\textbf{A}_{2}\textbf{x}_{1} + \textbf{B}_{2}\textbf{x}_{1}^{*} ~\cdots ~\textbf{A}_{N_{t}}\textbf{x}_{1} + \textbf{B}_{N_{t}}\textbf{x}_{1}^{*}\right],
\end{equation*}
\begin{equation*}
\textbf{X}_{2} = \left[\textbf{A}_{1}\textbf{x}_{2} + \textbf{B}_{1}\textbf{x}_{2}^{*} ~\textbf{A}_{2}\textbf{x}_{2} + \textbf{B}_{2}\textbf{x}_{2}^{*} ~\cdots ~\textbf{A}_{N_{t}}\textbf{x}_{2} + \textbf{B}_{N_{t}}\textbf{x}_{2}^{*}\right] 
\end{equation*}
where 
\begin{equation*}
\left\lbrace \textbf{A}_{i}, \textbf{B}_{i}  \in \mathbb{C}^{T \times k} ~|~ i = 1 \mbox{ to } N_{t}\right\rbrace 
\end{equation*}
is the set of column vector representation matrices \cite{Xl} of $\textbf{X}_{1}$, $\textbf{X}_{2}$ and $\textbf{x}_{1}^{T} = \left[x_{11} ~x_{12} ~\cdots ~x_{1k} \right]$, $\textbf{x}_{2}^{T} = \left[x_{21} ~x_{22} ~\cdots ~x_{2k} \right]$. If the above two designs are employed in a two-user MIMO-MAC, the vector received at the $j$-th antenna of the destination is of the form,
\begin{equation*}
\textbf{Y}_{c}(j) = \sqrt{\frac{\rho}{2N_{t}}}\textbf{X}_{1}\textbf{H}_{c}^{(1)}(j) +  \sqrt{\frac{\rho}{2N_{t}}}\textbf{X}_{2}\textbf{H}_{c}^{(2)}(j) + \textbf{N}_{c}(j).
\end{equation*}
Throughout the paper, it is assumed that the destination performs sphere decoding for the symbols of User-1 and User-2 jointly. Therefore, the complex variables of the two designs need to take values from a lattice constellation and hence square $M$-QAM constellation is used as the underlying constellation in the rest of this paper. Also, the channel equation has to be rewritten in a particular form in real variables which is amenable for sphere decoding. Towards that direction, using the column vector representations of $\textbf{X}_{1}$ and $\textbf{X}_{2}$, for each $j = 1$ to $N_{r}$, $\textbf{Y}_{c}(j)$ can be written in terms of its real and imaginary components as,
\begin{equation}
\label{real_s_antenna_mimo_mac_channel}
\vec{\textbf{Y}}_{c}(j) = \sqrt{\frac{\rho}{2N_{t}}}\tilde{\textbf{H}}_{c}^{(1)}(j)\vec{\textbf{x}}_{1} + \sqrt{\frac{\rho}{2N_{t}}}\tilde{\textbf{H}}_{c}^{(2)}(j)\vec{\textbf{x}}_{2} + \vec{\textbf{N}}_{c}(j)
\end{equation}
where the matrices $\tilde{\textbf{H}}_{c}^{(1)}(j) \in \mathbb{R}^{2T \times 2k}$ and $\tilde{\textbf{H}}_{c}^{(2)}(j) \in \mathbb{R}^{2T \times 2k}$ are as given in \eqref{L_G_user_1} and \eqref{L_G_user_2} respectively with $h^{(1)}_{i,j}$ denoting the channel from the $i$-th antenna of User-1 to the $j$-th antenna of the destination and $h^{(2)}_{i,j}$ denoting the channel from the $i$-th antenna of User-2 to the $j$-th antenna of the destination. 
\begin{figure*}
\begin{equation}
\label{L_G_user_1}
\tilde{\textbf{H}}_{c}^{(1)}(j) = \sum_{i = 1}^{N_{t}} \left[\begin{array}{cccc}
\mbox{Re}(h^{(1)}_{i,j}\textbf{A}_{i}) + \mbox{Re}(h^{(1)}_{i,j}\textbf{B}_{i}) & -\mbox{Im}(h^{(1)}_{i,j}\textbf{A}_{i}) + \mbox{Im}(h^{(1)}_{i,j}\textbf{B}_{i})\\
\mbox{Im}(h^{(1)}_{i,j}\textbf{A}_{i}) + \mbox{Im}(h^{(1)}_{i,j}\textbf{B}_{i}) & \mbox{Re}(h^{(1)}_{i,j}\textbf{A}_{i}) - \mbox{Re}(h^{(1)}_{i,j}\textbf{B}_{i})\\
\end{array}\right]
\end{equation}
\begin{equation}
\label{L_G_user_2}
\tilde{\textbf{H}}_{c}^{(2)}(j) = \sum_{i = 1}^{N_{t}} \left[\begin{array}{cccc}
\mbox{Re}(h^{(2)}_{i,j}\textbf{A}_{i}) + \mbox{Re}(h^{(2)}_{i,j}\textbf{B}_{i}) & -\mbox{Im}(h^{(2)}_{i,j}\textbf{A}_{i}) + \mbox{Im}(h^{(2)}_{i,j}\textbf{B}_{i})\\
\mbox{Im}(h^{(2)}_{i,j}\textbf{A}_{i}) + \mbox{Im}(h^{(2)}_{i,j}\textbf{B}_{i}) & \mbox{Re}(h^{(2)}_{i,j}\textbf{A}_{i}) - \mbox{Re}(h^{(2)}_{i,j}\textbf{B}_{i})\\
\end{array}\right]
\end{equation}
\hrule
\end{figure*}
Equation \eqref{real_s_antenna_mimo_mac_channel} can be written as
\begin{equation*}
\vec{\textbf{Y}}_{c}(j) = \sqrt{\frac{\rho}{2N_{t}}}\left[\tilde{\textbf{H}}^{1}_{c}(j) ~\tilde{\textbf{H}}^{2}_{c}(j)\right] \textbf{z} + \vec{\textbf{N}}_{c}(j)
\end{equation*}
where $\textbf{z} = \left[ (\vec{\textbf{x}}_{1})^{T} ~(\vec{\textbf{x}}_{2})^{T} \right]^{T} \in \mathbb{R}^{4k \times 1}$. Juxtaposing $\vec{\textbf{Y}}_{c}(j)$ for all $j = 1$ to $N_{r}$ one below the other, the channel equation is given by
\begin{equation}
\label{real_channel_equation}
\textbf{y} = \sqrt{\frac{\rho}{2N_{t}}}\textbf{M}\textbf{z} + \textbf{n}
\end{equation}
where
\begin{equation*}
\textbf{y} = \left[ (\vec{\textbf{Y}}_{c}(1))^{T} ~ (\vec{\textbf{Y}}_{c}(2))^{T} \cdots (\vec{\textbf{Y}}_{c}(N_{r}))^{T} \right]^{T} \in \mathbb{R}^{2TN_{r} \times 1},
\end{equation*}
\begin{equation*}
\textbf{n} = \left[ (\vec{\textbf{N}}_{c}(1))^{T} ~ (\vec{\textbf{N}}_{c}(2))^{T} \cdots (\vec{\textbf{N}}_{c}(N_{r}))^{T} \right]^{T} \in \mathbb{R}^{2TN_{r} \times 1}
\end{equation*}
and
\begin{equation}
\label{M_matrix}
\textbf{M} = \left[\begin{array}{cccc}
\tilde{\textbf{H}}_{c}^{(1)}(1) & \tilde{\textbf{H}}_{c}^{(2)}(1)\\
\tilde{\textbf{H}}_{c}^{(1)}(2)  & \tilde{\textbf{H}}_{c}^{(2)}(2)\\
\vdots & \vdots\\
\tilde{\textbf{H}}_{c}^{(1)}(N_{r}) & \tilde{\textbf{H}}_{c}^{(2)}(N_{r})\\
\end{array}\right] \in \mathbb{R}^{2TN_{r} \times 4k}.
\end{equation}
The matrix $\textbf{M}$ can be used as the lattice generator for carrying out sphere decoding algorithm. Since the variables of the two designs take values from an identical square $M$-QAM constellation, each component of $\textbf{z}$ takes value from the corresponding $\sqrt{M}$-PAM signal set. For $\textbf{M}$ to have rank $4k$, the inequality $2TN_{r} \geq 4k$ must hold. Hence, throughout the paper, we assume $N_{r} = \lceil\frac{2k}{T}\rceil$. Viewing the lattice generator $\textbf{M}$ as a real linear design in the variables $\mbox{Re}(h^{(1)}_{i,j})$, $\mbox{Im}(h^{(1)}_{i,j})$, $\mbox{Re}(h^{(2)}_{i,j})$ and $\mbox{Im}(h^{(2)}_{i,j})$, $\textbf{M}$ can also be written using the column vector representation as shown below,
\begin{equation*}
\textbf{M} = \left[\textbf{C}_{1}\textbf{h}^{(1)} ~\textbf{C}_{2}\textbf{h}^{(1)} \cdots \textbf{C}_{2k}\textbf{h}^{(1)} ~\textbf{C}_{1}\textbf{h}^{(2)} ~\textbf{C}_{2}\textbf{h}^{(2)} \cdots \textbf{C}_{2k}\textbf{h}^{(2)}\right] 
\end{equation*}
where

{\small
\begin{equation*}
\textbf{h}^{(1)} = \left[ (\vec{\textbf{H}}_{c}^{(1)}(1))^{T} ~ (\vec{\textbf{H}}_{c}^{(1)}(2))^{T} \cdots (\vec{\textbf{H}}_{c}^{(1)}(N_{r}))^{T} \right]^{T} \in \mathbb{R}^{2N_{t}N_{r} \times 1},
\end{equation*}
\begin{equation*}
\textbf{h}^{(2)} = \left[ (\vec{\textbf{H}}_{c}^{(2)}(1))^{T} ~ (\vec{\textbf{H}}_{c}^{(2)}(2))^{T} \cdots (\vec{\textbf{H}}_{c}^{(2)}(N_{r}))^{T} \right]^{T} \in \mathbb{R}^{2N_{t}N_{r} \times 1},
\end{equation*}
}

\noindent and $\left\lbrace \textbf{C}_{i} \in \mathbb{R}^{2TN_{r} \times 2N_{t}N_{r}}~|~ i = 1 \mbox{ to } 2k\right\rbrace$
is the set of column vector representation matrices of $\textbf{M}$. Since the design employed for both the users is the same, observe that the set of column vector representation matrices for the first $2k$ columns of $\textbf{M}$ and the last $2k$ columns of $\textbf{M}$ are the same.
\begin{definition}
\label{def_1}
\cite{YiK} A matrix is said to be column (row) monomial, if there is at most one non-zero entry in every column (row) of it. 
\end{definition}

\indent In this paper, we design a special class of complex designs such that the resulting $\textbf{M}$ has the following properties:
\begin{itemize}
\item (p.1). The entries in the first $2k$ columns of $\textbf{M}$ are of the form $\pm \mbox{Re}(h^{(1)}_{i,j})$, $\pm \mbox{Im}(h^{(1)}_{i,j})$ $\forall ~i, j$.
\item (p.2). The entries in the last $2k$ columns of $\textbf{M}$ are of the form $\pm \mbox{Re}(h^{(2)}_{i,j})$, $\pm \mbox{Im}(h^{(2)}_{i,j}) ~\forall ~i, j$.
\item (p.3). Every column of $\textbf{M}$ has all the $2N_{t}N_{r}$ variables appearing exactly once.
\end{itemize}
The above three properties imply that for each $i = 1$ to $2k, \textbf{C}_{i}$ is both row and column monomial. The class of lattice generators with the above set of conditions are referred as row-column monomial lattice generators which are formally defined as below.
\begin{definition}
\label{def_2}
A lattice generator, $\textbf{M}$ is said to be row-column monomial (RC monomial) if the column vector representation matrices of $\textbf{M}$ are both row and column monomial.
\end{definition}

\indent Note that the property (p.3) implies that the norm of the first $2k$ columns of $\textbf{M}$ are equal. Similarly, the norm of the last $2k$ columns of $\textbf{M}$ are equal.
\section{Structure on $\textbf{M}$ for reduction in the decoding complexity}
\label{sec3}
Applying Q-R decomposition on $\textbf{M}$ and multiplying $\textbf{Q}^{T}$ on both the sides of the channel equation in \eqref{real_channel_equation}, we have
\begin{equation}
\label{q_real_channel_equation}
\tilde{\textbf{y}} = \sqrt{\frac{\rho}{2N_{t}}}\textbf{R}\textbf{z} + \tilde{\textbf{n}}
\end{equation}
where $\tilde{\textbf{y}} = \textbf{Q}^{T}\textbf{y} \in \mathbb{R}^{2TN_{r} \times 1}, \tilde{\textbf{n}} = \textbf{Q}^{T}\textbf{n} \in \mathbb{R}^{2TN_{r} \times 1}$
and $\textbf{R} \in \mathbb{R}^{2TN_{r} \times 4k}$. Since we have assumed $N_{r} = \lceil\frac{2k}{T}\rceil$, only the first 4k rows of $\textbf{R}$ have non-zero entries and hence $\tilde{\textbf{y}}$ is essentially a $4k \times 1$ vector and $\textbf{R}$ is essentially a square matrix (neglecting the last $2TN_{r} - 4k$ rows) given by, 
\begin{equation*}
\textbf{R} = \left[\begin{array}{cccc}
\textbf{R}_{1,1} & \textbf{R}_{1,2}\\
\textbf{0} & \textbf{R}_{2,2}\\
\end{array}\right]
\end{equation*}
with $\textbf{R}_{1,1}, \textbf{R}_{1, 2}, \textbf{R}_{2, 2} \in \mathbb{R}^{2k \times 2k}$ such that $\textbf{R}_{1,1}$ and $\textbf{R}_{2, 2}$ are upper triangular matrices. The ML decoding metric is given by
\begin{equation}
\label{ml_decoder}
\hat{\textbf{z}} = \mbox{arg} \min_{\textbf{z}}||\tilde{\textbf{y}} - \sqrt{\frac{\alpha}{2N_{t}}}\textbf{R}\textbf{z}||^{2}.
\end{equation}

The following proposition shows that the entries in the sub-matrix $\textbf{R}_{1,2}$ cannot be made zero when identical STBCs are employed in the two-user MIMO-MAC set-up.

\begin{proposition}
\label{thm_1}
When identical STBCs are employed in a two-user MIMO-MAC, it is not possible to have zero entries in the matrix $\textbf{R}_{1,2}$.
\end{proposition}
\begin{proof}
The matrix $\textbf{R}$ arising out of the Q-R decomposition of $\textbf{M}$ is of the form,
\begin{equation}
\label{R_matrix}
\textbf{R} =\left[\begin{array}{ccccc}
\langle\textbf{e}_{1}, \textbf{c}_{1}\rangle & \langle\textbf{e}_{1}, \textbf{c}_{2}\rangle & \cdots & \langle\textbf{e}_{1}, \textbf{c}_{4k}\rangle\\
0 & \langle\textbf{e}_{2}, \textbf{c}_{2}\rangle & \cdots & \langle\textbf{e}_{2}, \textbf{c}_{4k}\rangle \\
0 & 0 & \cdots & \langle\textbf{e}_{3}, \textbf{c}_{4k}\rangle\\
\vdots & \vdots & \ddots & \vdots\\
0 & 0 & 0 & \langle\textbf{e}_{4k}, \textbf{c}_{4k}\rangle
\end{array}\right]
\end{equation}
where $\textbf{c}_{i}$ denotes the $i$-th column of $\textbf{M}$, $\textbf{e}_{i} = \frac{\textbf{u}_{i}}{|\textbf{u}_{i}|}$ with
\begin{equation}
\label{expand_u}
\textbf{u}_{i} = \textbf{c}_{i} - \sum_{j = 1}^{i-1} \langle\textbf{e}_{j}, \textbf{c}_{i}\rangle\textbf{e}_{j} ~\forall i = 1 \cdots 4k.
\end{equation}
Note that for $1 \leq m \leq 2k$ and $2k +1 \leq n \leq 4k$, $\left[\textbf{R}\right]_{m, n}$ is given by,
\begin{equation*}
\left[\textbf{R}\right]_{m, n} = \langle\textbf{e}_{m}, \textbf{c}_{n}\rangle.
\end{equation*}
Also, note that the variables in the first $2k$ columns of $\textbf{M}$ do not appear in the last $2k$ columns of $\textbf{M}$. In particular, $\textbf{e}_{m}$ is a vector in the variables $\mbox{Re}(h^{(1)}_{i,j}), \mbox{Im}(h^{(1)}_{i,j})$ whereas $\textbf{c}_{m}$ is a vector in the variables $\mbox{Re}(h^{(2)}_{i,j}), \mbox{Im}(h^{(2)}_{i,j})$. Therefore, for any STBC employed in a two-user MIMO-MAC, the matrix $\textbf{R}_{1,2}$ cannot have zero entries unless there exists at least one pair of columns (say $\textbf{c}_{i}$ and $\textbf{c}_{j}$) in the first $2k$ columns of $\textbf{M}$ which are orthogonal.
\end{proof}

From the above proposition, constructing STBCs which give rise to both $\textbf{R}_{1,1} \in \mathcal{D}$ and $\textbf{R}_{2, 2} \in \mathcal{D}$ is the best thing that can be done towards constructing STBCs with reduced SDC. Hence, we study STBCs from a special class of complex designs which results in $\textbf{M}$ (through \eqref{M_matrix}) such that the Q-R decomposition of $\textbf{M}$ gives rise to the $\textbf{R}$ matrix with (i)
$\textbf{R}_{1,1} \in \mathcal{D} \mbox{ and } \textbf{R}_{2, 2} \in \mathcal{D}$ and (ii) $\textbf{R}_{1,1} \in \mathcal{D} \mbox{ and } \textbf{R}_{2, 2} \notin \mathcal{D}$ (such classes of STBCs are formally defined below).\\

\begin{definition}
\label{def_3}
For a two-user MIMO-MAC, an STBC is said to have reduced average SDC (ASDC), if the corresponding $\textbf{R}$ matrix is such that both $\textbf{R}_{1,1}, \textbf{R}_{2, 2} \in \mathcal{D}$.\\
\end{definition}

\begin{definition}
\label{def_4}
For a two-user MIMO-MAC, an STBC is said to have reduced worst-case SDC (WSDC), if the corresponding $\textbf{R}$ matrix is such that only $\textbf{R}_{1,1} \in \mathcal{D}$ (but $\textbf{R}_{2,2} \notin \mathcal{D}$).\\
\end{definition}

\indent In the next subsection, we quantify the reduction in the decoding complexity when both $\textbf{R}_{1,1}$ and $\textbf{R}_{2, 2}$ are diagonal matrices.
\subsection{Reduction in the decoding complexity when $\textbf{R}_{1,1}, \textbf{R}_{2,2} \in \mathcal{D}$}
\label{subsec1_sec3}
For the decoder given by \eqref{ml_decoder}, we quantify the reduction in the decoding complexity when $\textbf{R}_{1,1}, \textbf{R}_{2,2} \in \mathcal{D}$. For point to point co-located MIMO channels, the SDC has been reduced in \cite{BHV}, \cite{MoB} and \cite{PaR} by making certain entries of $\textbf{R}$ matrix take zero values. Since $\textbf{R}$ is upper triangular, the ML decoding metric in \eqref{ml_decoder} can be split as
\begin{equation*}
||\tilde{\textbf{y}}_{1} - \sqrt{\frac{\alpha}{2N_{t}}}(\textbf{R}_{1,1}\vec{\textbf{x}}_{1} + \textbf{R}_{1,2}\vec{\textbf{x}}_{2})||^{2} + ||\tilde{\textbf{y}}_{2} - \sqrt{\frac{\alpha}{2N_{t}}}\textbf{R}_{2,2}\vec{\textbf{x}}_{2}||^{2}
\end{equation*}
where 
\begin{equation*}
\tilde{\textbf{y}}_{1} = \left[\tilde{\textbf{y}}(1) ~\tilde{\textbf{y}}(2) \cdots \tilde{\textbf{y}}(2k) \right]^{T}
\end{equation*}
and
\begin{equation*}
\tilde{\textbf{y}}_{2} = \left[\tilde{\textbf{y}}(2k+1) ~\tilde{\textbf{y}}(2k+2) \cdots \tilde{\textbf{y}}(4k) \right]^{T}.
\end{equation*}
Note that each component of $\vec{\textbf{x}}_{2}$ takes value from $\sqrt{M}$-PAM and hence the vector $\vec{\textbf{x}}_{2}$ totally takes $M^{k}$ distinct values. For a particular choice of $\vec{\textbf{x}}_{2}$, say $\vec{\textbf{x}}_{2} = \textbf{a}$, the metric for decoding $\vec{\textbf{x}}_{1}$ is
\begin{equation}
||\tilde{\textbf{y}}_{1}^{a} - \sqrt{\frac{\alpha}{2N_{t}}}\textbf{R}_{1,1}\vec{\textbf{x}}_{1}||^{2} + ||\tilde{\textbf{y}}_{2}^{a}||^{2}
\end{equation}
where
\begin{equation*}
\tilde{\textbf{y}}_{1}^{a} = \tilde{\textbf{y}}_{1} - \sqrt{\frac{\alpha}{2N_{t}}}\textbf{R}_{1,2}\textbf{a} \mbox{  and }\tilde{\textbf{y}}_{2}^{a} = \tilde{\textbf{y}}_{2} - \sqrt{\frac{\alpha}{2N_{t}}}\textbf{R}_{2,2}\textbf{a}.
\end{equation*}
Since $\textbf{R}_{1,1} \in \mathcal{D}$, for each $i$ = $1$ to $2k$, the $i$-th real variable of $\vec{\textbf{x}}_{1}$ can be decoded independent of the other real variables as 
\begin{equation*}
\hat{\vec{\textbf{x}}}_{1}(i) = \mathcal{Q}\left( \frac{\tilde{\textbf{y}}_{1}^{a}(i)}{\sqrt{\frac{\alpha}{2N_{t}}}\left[\textbf{R}_{1,1}\right]_{i,i}}\right) 
\end{equation*}
where $\mathcal{Q}(.)$ denotes the nearest integer quantizer operation whose complexity is independent of the size of the constellation. Therefore, the worst case decoding complexity is $O(M^{2k})$. Note that, the worst case complexity of the decoder remains to be $O(M^{2k})$ irrespective of whether $\textbf{R}_{2,2} \in \mathcal{D}$ or otherwise. However, when $\textbf{R}_{2,2} \in \mathcal{D}$, the ASDC is reduced as follows: When $\textbf{R}_{2,2} \in \mathcal{D}$, in choosing a particular value for $\vec{\textbf{x}}_{2}$, $2k$ independent sorting operations are needed where each sorting operation involves sorting of $\sqrt{M}$ integers based on a constraint function. However, in the worst case, if $\textbf{R}_{2,2}$ is not diagonal (with all the upper diagonal entries of $\textbf{R}_{2,2}$ being nonzero), then there needs to be a single sorting operation of $M^{k}$ vectors of length $2k$ based on a constraint function. Thus, with $\textbf{R}_{2,2} \in \mathcal{D}$, there is a reduction in the sorting complexity which is significant especially when $M$ is large.

\subsection{Necessary and sufficient conditions on $\textbf{M}$ such that $\textbf{R}_{1,1}$, $\textbf{R}_{2,2} \in \mathcal{D}$ }
\label{subsec2_sec3}
\indent In this subsection, a set of necessary and sufficient conditions on the matrix set $\left\lbrace \textbf{C}_{1}, \textbf{C}_{2} \cdots \textbf{C}_{2k} \right\rbrace$ is provided such that both $\textbf{R}_{1,1}$ and $\textbf{R}_{2, 2}$ are diagonal matrices. The following definition is important towards proving the necessary and sufficient conditions.

\begin{definition}
\label{def_5}
A $k$-group partition of the index set $\mathcal{I}_{2k} = \left\lbrace 1, 2 \cdots 2k\right\rbrace$ consists of $k$ disjoint subsets, $\mathcal{G}_{1}, \mathcal{G}_{2}, \cdots \mathcal{G}_{k}$ such that $|\mathcal{G}_{i}| = 2 ~\forall ~i = 1 \mbox{ to }k$.
\end{definition}

\begin{theorem}
\label{thm_2}
The Q-R decomposition of $\textbf{M}$ results in a $\textbf{R}$ matrix with $\textbf{R}_{1,1}, \textbf{R}_{2,2} \in \mathcal{D}$ if and only if the matrix set $\left\lbrace \textbf{C}_{1}, \textbf{C}_{2} \cdots \textbf{C}_{2k} \right\rbrace$ satisfies the following conditions:
\begin{enumerate}
\item For $i \neq j$, the matrices in the set $\left\lbrace \textbf{C}_{1}, \textbf{C}_{2} \cdots \textbf{C}_{2k} \right\rbrace$ must be Hurwitz-Radon orthogonal, i.e.,
\begin{equation*}
\textbf{C}_{i}^{T}\textbf{C}_{j} + \textbf{C}_{j}^{T}\textbf{C}_{i} = \textbf{0}_{2N_{t}N_{r}}.
\end{equation*}
\item For a fixed $l, m \in \mathcal{I}_{2k}$ such that $l \neq m$, there exists a $k$-group partition of $\mathcal{I}_{2k}$ given by $\mathcal{P}^{l,m} = \left\lbrace \mathcal{G}_{1}^{l,m}, \mathcal{G}_{2}^{l,m}, \cdots \mathcal{G}_{k}^{l,m}\right\rbrace$
such that 
\begin{equation*}
\textbf{C}_{l}^{T}\textbf{C}_{\mathcal{G}_{i}^{l,m}(1)} =  \textbf{C}_{m}^{T}\textbf{C}_{\mathcal{G}_{i}^{l,m}(2)} \mbox{ and }
\end{equation*} 
\begin{equation*}
\textbf{C}_{m}^{T}\textbf{C}_{\mathcal{G}_{i}^{l,m}(1)} =  -\textbf{C}_{l}^{T}\textbf{C}_{\mathcal{G}_{i}^{l,m}(2)} ~\forall i = 1 \mbox{ to } k.\\
\end{equation*}
\end{enumerate}
\end{theorem}
\begin{proof}
The 'if' part can be proved by substituting the conditions 1) and 2) (given in the statement of the theorem) in $\textbf{R}$ which is straightforward. Hence, we prove the 'only if' part of the theorem. Since $\textbf{R}_{1,1} \in \mathcal{D}$, $ \langle\textbf{e}_{i}, \textbf{c}_{j}\rangle = 0$ for all $i \neq j$ such that $1 \leq i, j \leq 2k$. This implies $ \langle\textbf{c}_{i}, \textbf{c}_{j}\rangle  = 0$ for all $i \neq j$ such that $1 \leq i, j \leq 2k$. Therefore, the first $2k$ columns of $\textbf{M}$ are necessarily orthogonal to each other and hence	
\begin{equation*}
\textbf{C}_{i}^{T}\textbf{C}_{j} + \textbf{C}_{j}^{T}\textbf{C}_{i} = \textbf{0}_{2N_{t}N_{r}} \mbox{ for all } i \neq j.\\
\end{equation*}
This proves the condition 1) of the theorem (This condition reduces the WSDC). In the rest of the proof, the condition in 2) is proved. \\
\indent The structure of the matrix $\textbf{R}_{2,2}$ is given in \eqref{R_2_2} (shown at the top of the next page).
\begin{figure*}
\begin{equation}
\label{R_2_2}
\textbf{R}_{2,2} =\left[\begin{array}{ccccc}
\langle\textbf{e}_{2k + 1}, \textbf{c}_{2k + 1}\rangle & \langle\textbf{e}_{2k + 1}, \textbf{c}_{2k + 2}\rangle & \cdots & \langle\textbf{e}_{2k + 1}, \textbf{c}_{4k}\rangle\\
0 & \langle\textbf{e}_{2k + 2}, \textbf{c}_{2k + 2}\rangle & \cdots & \langle\textbf{e}_{2k + 2}, \textbf{c}_{4k}\rangle \\
0 & 0 & \cdots & \langle\textbf{e}_{2k + 3}, \textbf{c}_{4k}\rangle\\
\vdots & \vdots & \ddots & \vdots\\
0 & 0 & 0 & \langle\textbf{e}_{4k}, \textbf{c}_{4k}\rangle
\end{array}\right]
\end{equation}
\hrule
\end{figure*}
Since $\textbf{R}_{2,2} \in \mathcal{D}$, we have $\langle\textbf{e}_{l}, \textbf{c}_{m}\rangle = 0$ for $l \neq m$ such that $2k + 1 \leq l, m \leq 4k$. This implies 
\begin{equation*}
\langle\textbf{u}_{l}, \textbf{c}_{m}\rangle = 0.
\end{equation*}
Using \eqref{expand_u} in the above equation, we have
\begin{equation*}
\langle\textbf{u}_{l}, \textbf{c}_{m}\rangle =  \langle\textbf{c}_{l}, \textbf{c}_{m}\rangle  - \sum_{j = 1}^{l-1} \langle\textbf{e}_{j}, \textbf{c}_{l}\rangle\langle\textbf{e}_{j}, \textbf{c}_{m}\rangle = 0.
\end{equation*}
Since $\textbf{C}_{i}^{T}\textbf{C}_{j} + \textbf{C}_{j}^{T}\textbf{C}_{i} = \textbf{0}_{2N_{t}N_{r}}$ for all $i \neq j$, we have $\langle\textbf{c}_{l}, \textbf{c}_{m}\rangle = 0$ and hence
\begin{equation*}
 \sum_{j = 1}^{l-1} \langle\textbf{e}_{j}, \textbf{c}_{l}\rangle\langle\textbf{e}_{j}, \textbf{c}_{m}\rangle = 0. 
\end{equation*}
As $l$ takes value from $2k + 1$ to $4k$, the above summation can be split as  
\begin{equation*}
 \sum_{j = 1}^{2k} \langle\textbf{e}_{j}, \textbf{c}_{l}\rangle\langle\textbf{e}_{j}, \textbf{c}_{m}\rangle +  \sum_{j = 2k + 1}^{l-1} \langle\textbf{e}_{j}, \textbf{c}_{l}\rangle\langle\textbf{e}_{j}, \textbf{c}_{m}\rangle= 0. 
\end{equation*}
Since $\textbf{R}_{2,2} \in \mathcal{D}$, each term in the second summand of the above equation is individually zero and hence we have
\begin{equation*}
\sum_{j = 1}^{2k} \langle\textbf{e}_{j}, \textbf{c}_{l}\rangle\langle\textbf{e}_{j}, \textbf{c}_{m}\rangle = 0.
\end{equation*}
As the first $2k$ columns of $\textbf{M}$ are orthogonal to each other and have equal norms, we have
\begin{equation*}
\sum_{j = 1}^{2k} \langle\textbf{c}_{j}, \textbf{c}_{l}\rangle\langle\textbf{c}_{j}, \textbf{c}_{m}\rangle = 0.
\end{equation*}
As $2k + 1 \leq l, m \leq 4k$, we have $\textbf{c}_{l} = \textbf{C}_{l'}\textbf{h}^{(2)}$, $\textbf{c}_{m} = \textbf{C}_{m'}\textbf{h}^{(2)}$ and $\textbf{c}_{j} = \textbf{C}_{j}\textbf{h}^{(1)}$ where $l' = l~\mbox{modulo}~2k$ and $m' = m~\mbox{modulo}~2k$ and hence
\begin{equation*}
\sum_{j = 1}^{2k} ((\textbf{h}^{(2)})^{T}\textbf{C}_{l}^{T}\textbf{C}_{j}\textbf{h}^{(1)})((\textbf{h}^{(2)})^{T}\textbf{C}_{m}^{T}\textbf{C}_{j}\textbf{h}^{(1)}) = 0.
\end{equation*}
Note that for a fixed $m$, the matrices $\textbf{C}_{m}^{T}\textbf{C}_{j}$ and $\textbf{C}_{m}^{T}\textbf{C}_{i}$ do not have nonzero entries at the same position for all $i \neq j$. Similarly, for a fixed $l$, the matrices $\textbf{C}_{l}^{T}\textbf{C}_{j}$ and $\textbf{C}_{l}^{T}\textbf{C}_{i}$ do not have nonzero entries at the same position for all $i \neq j$. Hence, for a given $l, m$, there exists a $k$-group partition  $\mathcal{P}^{l,m} = \left\lbrace \mathcal{G}_{1}^{l,m}, \mathcal{G}_{2}^{l,m}, \cdots \mathcal{G}_{k}^{l,m}\right\rbrace$ of the index set $\mathcal{I}_{2k}$ such that
\begin{equation*}
\textbf{C}_{l}^{T}\textbf{C}_{\mathcal{G}_{i}^{l,m}(1)} =  \textbf{C}_{m}^{T}\textbf{C}_{\mathcal{G}_{i}^{l,m}(2)} \mbox{ and }
\end{equation*} 
\begin{equation*}
\textbf{C}_{m}^{T}\textbf{C}_{\mathcal{G}_{i}^{l,m}(1)} =  -\textbf{C}_{l}^{T}\textbf{C}_{\mathcal{G}_{i}^{l,m}(2)} ~\forall i = 1 \mbox{ to } k.\\
\end{equation*}
This completes the proof.
\end{proof}

\indent In the following section, we present explicit constructions of STBCs which have (i) reduced ASDC and (ii) reduced WSDC.

\section{Code constructions}
\label{sec4}
In this section, complex designs which results in the $\textbf{R}$ matrix with (i) $\textbf{R}_{1,1}, \textbf{R}_{2,2} \in \mathcal{D}$ and (ii) only $\textbf{R}_{1,1} \in \mathcal{D}$ are presented. Henceforth, we denote a complex design for $N_{t}$ antennas in $k$ variables as $\textbf{X}\left(N_{t}, k \right)$. First, we construct complex designs which results in $\textbf{R}_{1,1}, \textbf{R}_{2,2} \in \mathcal{D}$. Construction of these designs has been divided in to four cases depending on the values of $N_{t}$ and $k$.

\begin{case}\label{case1} $N_{t} = 2a$ and $k = 2b$ ($a$ and $b$ are positive integers): In this case, the design is constructed in the following 3 steps.
\begin{itemize}
\item Step (i) : Let $\mathbf{\Omega}_{m}$ represent a $2 \times 2$ Alamouti design in complex variables $x_{2m+ 1}$, $x_{2m+2}$ for each $m =  0, 1, \cdots b - 1$ as given below,
\begin{equation*}
\mathbf{\Omega}_{m}  = \left[\begin{array}{rr}
x_{2m+1} & -x_{2m+2}^{*}\\
x_{2m+2}  & x_{2m+1}^{*}\\
\end{array}\right].
\end{equation*}

\item Step (ii) : Using  $\mathbf{\Omega}_{m}$, construct a $2a \times 2a$ matrix $\textbf{X}_{m}$ given by
\begin{equation*}
\textbf{X}_{m} = \mathbf{\Omega}_{m} \otimes \textbf{I}_{2}^{\otimes (a - 1)} \mbox{ for each } m =  0, 1, \cdots b - 1.
\end{equation*}

\item Step (iii) : Using $\mathbf{\Omega}_{m}$, $\textbf{X}\left(N_{t}, k\right)$ is constructed as
\begin{equation*}
\textbf{X}\left(N_{t}, k \right)  = \left[ \textbf{X}_{0}^{T} ~  \textbf{X}_{1}^{T} ~ \cdots ~ \textbf{X}_{b - 1}^{T} \right]^{T}.
\end{equation*}
\end{itemize}
\end{case}

\begin{case}\label{case2} $N_{t} = 2a$ and $k = 2b + 1$ : In this case, $\textbf{X}\left(N_{t}, k\right)$ is constructed in two steps as given below.
\begin{itemize}
\item Step (i) : Construct $\textbf{X}\left(N_{t}, 2b\right)$ as given in Case \ref{case1}.
\item Step (ii) : $\textbf{X}\left(N_{t}, k\right)$ = $\left[ \textbf{X}\left(N_{t}, 2b\right)^{T} x_{2b+1}\textbf{I}_{N_{t}}\right]^{T}$.
\end{itemize}
\end{case}

\begin{case}\label{case3} $N_{t} = 2a+1$ and $k = 2b$ : In this case, $\textbf{X}\left(N_{t}, k\right)$ are constructed in the following 2 steps.
\begin{itemize}
\item Step (i) : Construct $\textbf{X}\left(N_{t}+2, k\right)$ as given in Case \ref{case1}.
\item Step (ii) : Drop the last column of $\textbf{X}\left(N_{t}+2, k\right)$.
\end{itemize}
\end{case}

\begin{case}\label{case4} $N_{t} = 2a+1$ and $k = 2b + 1$ : In this case, $\textbf{X}\left(N_{t}, k\right)$ are constructed in the following 2 steps.
\begin{itemize}
\item Step (i) : Construct $\textbf{X}\left(N_{t}+2, k\right)$ as given in Case \ref{case2}.
\item Step (ii) : Drop the last column of $\textbf{X}\left(N_{t}+2, k\right).$\\
\end{itemize}
\end{case}

\indent The rate (in complex symbols per channel use) of the above proposed designs is at most $\frac{2}{N_{t}}$. Therefore, whenever STBCs with minimum ASDC are desired (with both $\textbf{R}_{1,1} \in \mathcal{D}$ and $\textbf{R}_{2,2} \in \mathcal{D}$), there is a substantial loss in the rate of transmission especially when $N_{t} > 2$. However, if reduction of WSDC is targeted, then constructing complex designs which lead to only $\textbf{R}_{1,1} \in \mathcal{D}$ is sufficient. The following theorem states that the class of complex orthogonal designs \cite{TJC}, \cite{Xl} (other than Alamouti design) results in the class of RC monomial lattice generators which in-turn lead to $\textbf{R}_{1,1} \in \mathcal{D}$ (but $\textbf{R}_{2,2} \notin \mathcal{D}$). 
\begin{theorem}
For $N_{t} > 2$, STBCs from square complex orthogonal designs (CODs) reduce the WSDC for a two-user MIMO-MAC.
\end{theorem}
\begin{proof}
We have to show that STBCs from the class of CODs (other than Alamouti design) results in a class of RC monomial lattice generators which in-turn lead to $\textbf{R}_{1,1} \in \mathcal{D}$ but $\textbf{R}_{2,2} \notin \mathcal{D}$. It is straightforward to verify that the column vector representation matrices $\left\lbrace \textbf{C}_{i} ~|~ i = 1 \mbox{ to } 2k \right\rbrace$ of $\textbf{M}$ arising from CODs satisfy the sufficient condition 1) of Theorem \ref{thm_2}. Hence, the corresponding class of $\textbf{R}$ matrices satisfy $\textbf{R}_{1,1} \in \mathcal{D}$.\\
\indent In the rest of this paragraph, we only provide a sketch of the proof to show that the matrices $\left\lbrace \textbf{C}_{i} ~|~ i = 1 \mbox{ to } 2k \right\rbrace$ arising from CODs do not satisfy the sufficient conditions in 2) of Theorem \ref{thm_2} (this is to prove that $\textbf{R}_{2,2} \notin \mathcal{D}$). Recall that a COD in $a + 1$ complex variables for $N_{t} = 2^{a}$ antennas can be constructed in a recursive fashion from a COD in $a$ variables for $N_{t} = 2^{a-1}$ antennas for all $a \geq 2$ (See Section III. D in \cite{Xl}). We use the recursive construction technique of CODs to prove our result. First, it can be shown that the matrices $\left\lbrace \textbf{C}_{i} ~|~ i = 1 \mbox{ to } 6 \right\rbrace$ arising from the COD for $N_{t} = 4$ antennas do not satisfy the sufficient condition 2) of Theorem \ref{thm_2}. Then, from the recursive construction technique of CODs, it can be proved that the matrices $\left\lbrace \textbf{C}_{i} ~|~ i = 1 \mbox{ to } 2k \right\rbrace$ arising CODs with larger number of antennas do not satisfy the sufficient conditions in 2) of Theorem \ref{thm_2} as well. This completes the proof.
\end{proof}

\indent From the above theorem, it is clear that when only the WSDC is to be reduced, the rate of transmission can be increased from $\frac{2}{N_{t}}$ to (i) $\frac{a+1}{2^a}$ for the case of square designs where $N_{t} = 2^{a}b$ for positive integers $a$ and $b$.
\section{Discussion}
\label{sec5}
In this paper, we have proposed STBCs with minimum SDC for a two-user MIMO-MAC. Some possible directions for further research are as follows: 
\begin{itemize}
\item The rate (in complex symbols per channel use) of the proposed class of STBCs (in Section \ref{sec4}) which reduces the ASDC is at most $\frac{2}{N_{t}}$ for each user. Using the necessary and sufficient conditions on the column vector representation matrices in Theorem \ref{thm_2}, upper bounds on the rate (in complex symbols per channel use) can be obtained and possibly STBCs with higher rates can be constructed. 
\item We have studied STBCs which result in a $\textbf{R}$ matrix such that $\textbf{R}_{1,1}$ and $\textbf{R}_{2,2}$ are diagonal matrices. Construction of STBCs which results in more number of non-zeros in the upper-diagonal entries of $\textbf{R}_{1,1}$ and $\textbf{R}_{2,2}$ is an interesting direction for future work. Such STBCs may have higher ASDC and/or higher WSDC but may lead to larger rates.
\end{itemize}
\section*{Acknowledgment}
This work was partly supported by the DRDO-IISc Program on Advanced Research in Mathematical Engineering through a research grant to B.S. Rajan.
\end{document}